\documentstyle[graphics,twocolumn]{mn}

\newcommand{\mincir}{\raise
-2.truept\hbox{\rlap{\hbox{$\sim$}}\raise5.truept 
\hbox{$<$}\ }}
\newcommand{\magcir}{\raise
-2.truept\hbox{\rlap{\hbox{$\sim$}}\raise5.truept
\hbox{$>$}\ }}
\newcommand{\minmag}{\raise-2.truept\hbox{\rlap{\hbox{$<$}}\raise
6.truept\hbox
{$>$}\ }}

\newcommand{\lya}{Lyman-$\alpha$~}

\newcommand{\be}{\begin{equation}}
\newcommand{\ee}{\end{equation}}
\newcommand{\ba}{\begin{eqnarray}}
\newcommand{\ea}{\end{eqnarray}}
\newcommand{\brr}{\begin{array}}
 
\newcommand{\err}{\end{array}}
\newcommand{\bc}{\begin{center}}
\newcommand{\ec}{\end{center}}

\newcommand{\mpch} {\rm $h^{-1}$ Mpc\,\,}
\newcommand{\gad}{\mbox{\small{GADGET-2}\,\,}}

\DeclareMathAlphabet{\mathsc}{OT1}{cmr}{m}{sc}
\def\testbx{bx}%
\DeclareRobustCommand{\ion}[2]{%
\relax\ifmmode
\ifx\testbx\f@series
{\mathbf{#1\,\mathsc{#2}}}\else
{\mathrm{#1\,\mathsc{#2}}}\fi
\else\textup{#1\,{\mdseries\textsc{#2}}}%
\fi}
\title[Testing the accuracy of the Hydro-PM approximation]{Testing the
accuracy of the Hydro-PM approximation in numerical simulations of 
the \lya forest}

\author[M. Viel, M.G. Haehnelt, V. Springel] {Matteo Viel$^{1}$,
Martin G. Haehnelt$^{1}$ \& Volker Springel$^{2}$ \\ $^1$ Institute of
Astronomy, Madingley Road, Cambridge CB3 0HA\\ $^2$
Max-Planck-Institut f\"ur Astrophysik, Karl-Schwarzschild-Str. 1,
Garching bei M\"unchen, Germany \\  }

\begin{document}

\maketitle
\begin{abstract}
  We implement the hydro-PM (HPM) technique (Gnedin \& Hui 1998) in the
  hydrodynamical simulation code \gad and quantify the differences between
  this approximate method and full hydrodynamical simulations of the \lya
  forest in a concordance $\Lambda$CDM model. At redshifts $z=3$ and $z=4$,
  the differences between the gas and dark matter (DM) distributions, as
  measured by the one-point distribution of density fluctuations, the density
  power spectrum and the flux power spectrum, systematically decrease with
  increasing resolution of the HPM simulation. However, reducing these
  differences to less than a few percent requires a significantly larger
  number of grid-cells than particles, with a correspondingly larger demand
  for memory.  Significant differences in the flux decrement distribution
  remain even for very high resolution hydro-PM simulations, particularly at
  low redshift. At $z=2$, the differences between the flux power spectra
  obtained from HPM simulations and full hydrodynamical simulations are
  generally large and of the order of 20-30\%, and do not decrease with
  increasing resolution of the HPM simulation.  This is due to the presence of
  large amounts of shock-heated gas, a situation which is not adequately
  modelled by the HPM approximation. We confirm the results of Gnedin \& Hui
  (1998) that the statistical properties of the flux distribution are
  discrepant by $\ga 5-20\%$ when compared to full hydrodynamical simulations.
  The discrepancies in the flux power spectrum are strongly scale- and
  redshift-dependent and extend to large scales.  Considerable caution is
  needed in attempts to use calibrated HPM simulations for quantitative
  predictions of the flux power spectrum and other statistical properties of
  the \lya forest.
\end{abstract}


\section{Introduction}

The prominent absorption features blue-ward of the \lya emission in the
spectra of high-redshift quasars (QSOs) are believed to arise from
smooth density fluctuations of a photoionised warm intergalactic medium
which trace the dark matter distribution in a relatively simple manner
(see Rauch 1998 and Weinberg et al. 1999 for reviews).  As a result,
the flux power spectrum of this `\lya forest' has become a powerful
quantitative probe of the matter power spectrum on scales of
$1\,h^{-1}$ to $40\,h^{-1}$ Mpc at redshifts $z=2-4$. At these scales
and redshifts, the matter distribution is linear or mildly non-linear,
a regime that can be accurately modelled with numerical
simulations. Such simulations have been used to obtain quantitative
estimates of the clustering amplitude and constraints on cosmological
parameters from the \lya forest (Croft et al. 1998; Croft et al.  1999;
McDonald et al. 2000; Hui et al. 2001; Croft et al. 2002; McDonald
2003; Viel et al. 2003; Viel, Haehnelt \& Springel 2004; Viel, Weller
\& Haehnelt 2004; McDonald et al.  2004a; Desjacques \& Nusser 2004) or
on astro-physical parameters (Theuns et al. 1998, Meiksin et
al. 2001, McDonald et al.  2004b, Bolton et al. 2005).

Unfortunately, the flux power spectrum does not only depend on the dark
matter (DM) distribution but also on the thermal state of the
intergalactic medium (IGM), and possibly on feedback effects due to
star formation and active galactic nuclei (AGN). Ideally, one would
like to use simulations which not only take into account the non-linear
gravitational clustering of the matter distribution but also all the
relevant hydrodynamics of the gas, including effects of galaxy
formation physics, such as radiative cooling and heating, star
formation and winds driven by stellar associations or AGN. However,
full hydrodynamical simulations of the \lya forest are computationally
very demanding. This makes their use for extensive parameter studies
difficult. In addition, some physical processes, such as the feedback
mechanisms, are still poorly understood.  Thus, the use of approximate
numerical calculations of the flux distribution of the \lya forest very
attractive, an approach that has been widely applied in previous work
(e.g. McGill 1990; Hui et al. 1997; Meiksin \& White 2001; Viel et
al. 2002b; Zhan et al. 2005).  Note that such approximate calculations
of the \lya flux distribution have been crucial in establishing the
modern paradigm for the origin of the \lya forest in the first place
(Bi 1993, Bi \& Davidsen 1997, Viel et al. 2002a).

In 1998, Gnedin \& Hui (GH) have proposed the `Hydrodynamic
Particle-Mesh method' (HPM) as an efficient numerical method to
approximate the formation and evolution of the \lya forest. This
technique is based on a particle-mesh (PM) approach for following the
evolution of dark matter. The gravitational potential of the PM solver
is then modified with an effective potential which mimics the effect of
gas pressure.  GH found that global statistical properties of the flux
distribution in HPM simulations are accurate to $\sim 5-20\%$ when
compared to full hydrodynamical simulations.  This prompted e.g.
McDonald et al.~(2004a) to use HPM simulations that were calibrated with
a small number of hydrodynamical simulations to obtain predictions of
the flux power spectrum for a wide range of cosmological and physical
parameters describing the thermal state of the gas.

The statistical errors of the flux power spectrum obtained from
high-resolution Echelle spectra are $\sim 4$~\% and can in principle
become as small as a few percent for large samples of low-resolution
spectra (e.g. Kim et al.~2004, McDonald et al. 2005). This has opened
up the exciting prospect to use the \lya forest to constrain
inflationary parameters and the nature of dark matter, based on high
accuracy measurements of the DM power spectrum inferred from the \lya
forest (Viel, Weller \& Hahenelt 2004; Seljak et al.  2004; Viel et al.
2005).  However, a prerequisite is the availability of accurate
predictions of the flux power spectrum for a wide range of parameters.

The hydrodynamical code \gad (Springel, Yoshida \& White 2001; Springel
2005), which we have used extensively in earlier work for full
hydrodynamical simulations of the \lya forest (Viel, Haehnelt \&
Springel 2004; Bolton et al.  2005), is a TreeSPH code which also
offers a PM algorithm which can optionally be used to calculate
long-range gravitational forces.  In this code the HPM method of GH can
therefore be easily implemented. This makes \gad well suited for a
detailed analysis of the accuracy and systematic uncertainties of the
HPM method by comparing simulations run with it to full hydrodynamical
TreeSPH-PM simulations.

In this paper, we perform such an analysis and investigate the
dependence of the discrepancies between HPM and full hydrodynamical
simulations on a range of numerical parameters for the relevant
redshift range $z=2-4$. Note that we here do not  intend to optimize 
the HPM method.

The outline of the paper is as follows. In
Section \ref{hydro}, we briefly describe the hydrodynamical code \gad
and we review the basic equations of the HPM formalism. We also show
that the HPM implementation of \gad and the HPM code by Gnedin \& Hui
give similar results for a suitable choice of numerical parameters.  In
Section \ref{compare2}, we discuss the differences between our HPM
implementation and full hydrodynamical simulation by analysing the
statistical properties of the flux distribution.  We further analyse the
effect of shock heating, the influence of various numerical parameters
on the results, and the CPU time and memory requirements.  Finally,
Section \ref{conclu} contains a summary and our conclusions.

\section{Simulation methods of the \lya forest}

\label{hydro}
\subsection{Full hydrodynamical simulations}

The hydrodynamical simulation code {\small GADGET-2} (Springel, Yoshida
\& White 2001; Springel 2005) can optionally employ a PM technique to
calculate long-range gravitational forces, resulting in a `TreePM'
scheme for gravitational forces.  We will use hydrodynamical
simulations run with this SPH/TreePM implementation of \gad as
``reference'' simulations to assess in detail the accuracy and
systematic uncertainties of the approximate HPM method. The TreePM
approach speeds up the calculation of long-range gravitational forces
considerably compared to a tree-only implementation.

All our simulations were performed with periodic boundary conditions
and an equal number of dark matter and gas particles. We employ the
`entropy-formulation' of SPH proposed by Springel \& Hernquist (2002).
Radiative cooling and heating processes are followed using an
implementation similar to that of Katz et al.~(1996) for a primordial
mix of hydrogen and helium. We have assumed a mean UV background
produced by quasars as given by Haardt \& Madau (1996), which leads to
reionisation of the Universe at $z\simeq 6$. The simulations are run
with heating rates increased by a factor of 3.3 in order to achieve
temperatures which are close to observed temperatures (Abel \& Haehnelt
1999, Schaye et al.  2000, Ricotti et al.  2000).

In order to maximise the speed of the dissipative hydrodynamical
simulations we have employed a simplified star-formation criterion in
the majority of our runs. All gas at densities larger than 1000 times
the mean density was turned into collisionless stars. The absorption
systems producing the \lya forest have small overdensity so this
criterion has little effect on flux statistics, while speeding up the
calculation by a factor of $\sim 6$, because the small dynamical times
that would otherwise arise in the highly overdense gas need not to be
followed.  In a pixel-to-pixel comparison with a simulation which
adopted the full multi-phase star formation model of Springel \&
Hernquist (2003) we explicitly checked for any differences introduced
by this approximation.  We found that the differences in the flux
probability distribution function were smaller than 2\%, while the
differences in the flux-power spectrum were smaller than 0.2 \%.  We
have also turned off all feedback options of {\small GADGET-2} in our
simulations. An extensive resolution and box size study has been
performed in Viel, Haehnelt \& Springel (2004) and in Bolton et
al. (2005).

For all simulations presented here we have adopted a box size of 30
comoving \mpch and the cosmological parameters $\Omega_{0{\rm m}}=
0.26$, $\Omega_{0\Lambda} = 0.74$, $\Omega_{0{\rm b}} = 0.0463$ and
$H_0=72\,{\rm km\,s^{-1}Mpc^{-1}}$, $\sigma_8=0.85$ and $n=0.95$ (the
parameters of the B2 simulation in Viel, Haehnelt \& Springel 2004).
The CDM transfer functions of all models have been taken from
Eisenstein \& Hu (1999).

\begin{figure*}
\center\resizebox{0.5\textwidth}{!}{\includegraphics{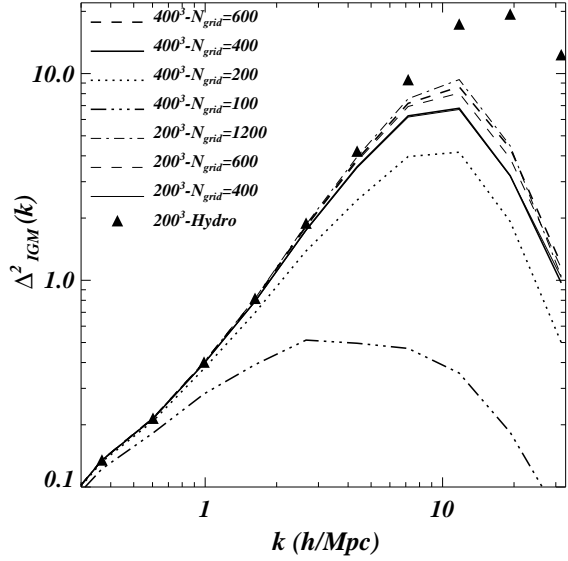}}
\caption{Power spectrum of the gas density field of the HPM simulations
run with \gad at $z=3$, at two different resolutions and for several
different values of the parameter $N_{\rm grid}$. The power spectrum of
the full hydrodynamical simulation is represented by the filled
triangles.}
\label{f1}
\end{figure*}

\begin{figure*}
\center\resizebox{1.0\textwidth}{!}{\includegraphics{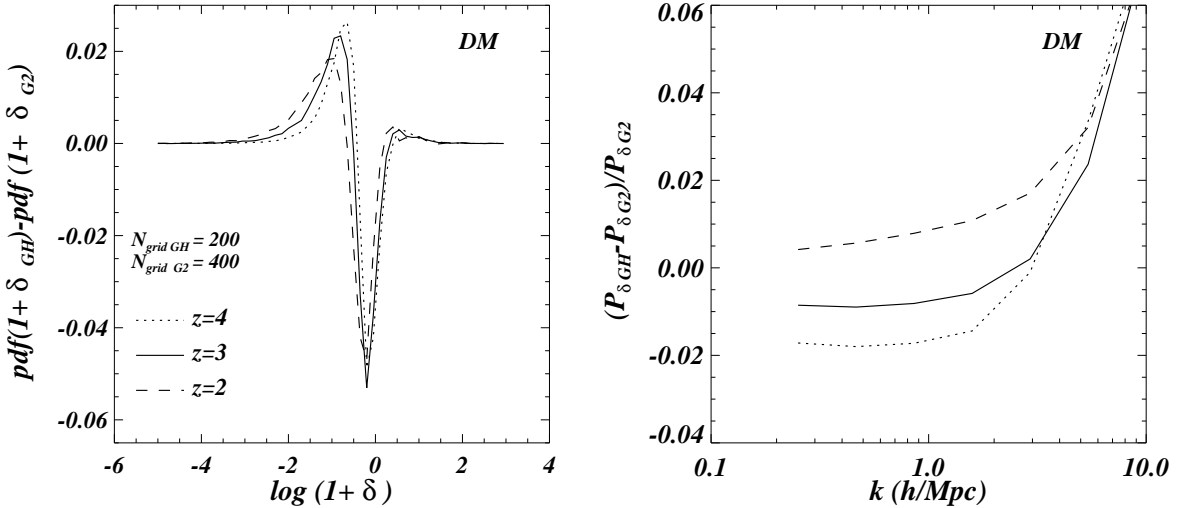}}
\caption{{\it Left:} differences in the probability distribution
functions of the dark matter density field between \gad (G2) and the Gnedin
\& Hui (GH) code. Both of them have been run in the PM mode with a grid
of $200^3$ for GH and $400^3$ for G2. {\it Right:} Fractional
differences in the 3D matter power spectrum. The results are shown at
three different redshifts $z=2,3,4$ as dashed, continuous and dotted
curves, respectively.}
\label{f2}
\end{figure*}

\begin{figure*}
\center\resizebox{1.0\textwidth}{!}{\includegraphics{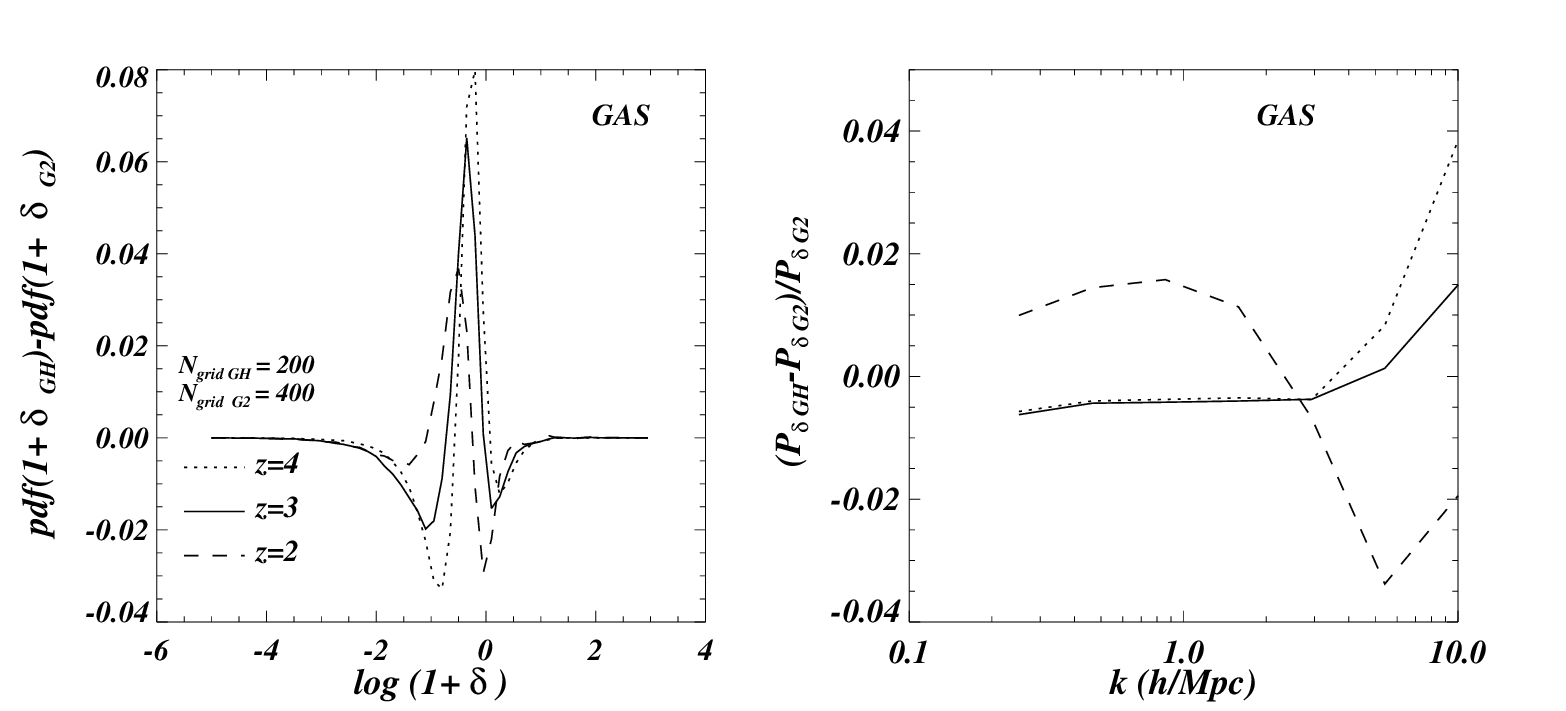}}
\caption{{\it Left:} differences in the probability distribution
functions of the gas  density field between simulations run with \gad
(G2) and the Gnedin
\& Hui (GH) code. Both of them have been run in the HPM mode with a grid
of $200^3$ for GH and $400^3$ for G2. {\it Right:} Fractional
differences in the 3D matter power spectrum. The results are shown at
three different redshifts $z=2,3,4$ as dashed, continuous and dotted
curves, respectively.}
\label{f3}
\end{figure*}

\subsection{HPM implementation of \gad}
\label{hpm}

GH proposed to introduce an effective potential that mimics gas pressure
into an otherwise collisionless dark matter simulation, carried out with a
particle mesh code. This method has become known as Hydro-Particle-Mesh (HPM)
approximation.  The idea of the HPM approximation is to take advantage of the
fact that the low density IGM responsible for most of the \lya forest
absorption obeys a simple relation between gas density and gas temperature,
which is well described by a power-law `equation of state': \begin{equation}
  T=T_0(z) \, (1+\delta)^{\gamma (z)-1}\;.
\label{eqst}
\end{equation}
The evolution of $T_0$ and $\gamma$ with redshift depends on the reionisation
history (Hui \& Gnedin 1997).  The `equation of state' predicts the
temperature of gas of given density to better than 10\% for the low density
IGM where shock heating is not important. Instead, the temperature is set by
a balance between photoionisation heating and adiabatic cooling due to the
expansion of the universe.

Based on the density alone, equation (\ref{eqst}) also allows an estimate of
the thermal pressure which enters the equation of motion for a cosmic gas
element.  We know from full hydrodynamical simulations that the baryons follow
the dark matter generally  well apart from high density regions where
pressure effects on small scales become important.  GH suggested therefore
to use the density of the dark matter in a PM simulation together with
Eqn.~(\ref{eqst}) to estimate the temperature and pressure of the gas. One can
then obtain the acceleration on a cosmic gas element due to the gradient of
the pressure as
\begin{equation}
        {d{\bmath v}\over dt} + H{\bmath v} = -\nabla\phi -
        {1\over\rho}\nabla P,
\label{eom}
\end{equation}
where ${\bmath v}$ is the gas peculiar velocity, $\phi$ is the gravitational
potential, and $P$ is the thermal pressure. If the gas is highly ionised (so
that the mean molecular weight is roughly constant, which is true for the
Lyman-alpha forest), and the temperature is a function of density only, so
that $P=P(\rho)$, equation (\ref{eom}) can be reduced to the expression
\begin{equation}
        {d{\bmath v}\over dt} + H{\bmath v} = -\nabla\psi ,
        \label{eompsi}
\end{equation}
where
\begin{equation}
        \psi = \phi + {\cal H},
        \label{psidef}
 \end{equation}
and ${\cal H}$, the {\it specific enthalpy\/}, is
\begin{equation}
        {\cal H}(\rho) = {P(\rho)\over\rho} + 
        \int_1^\rho {P(\rho^\prime)\over\rho^\prime}
        {d\rho^\prime\over\rho^\prime}.
\label{enthalpy}
\end{equation}

Equation (\ref{eompsi}) is identical to the equation of motion for the
collisionless dark matter except that the usual gravitational potential $\phi$
is replaced by an effective potential $\psi$, which takes into account both
gravity and thermal pressure.  Since the gravitational potential $\phi$ has to
be computed from the density field in a regular PM simulation anyway,
computing the enthalpy adds only a modest computational overhead.

\begin{figure*}
\center\resizebox{1.0\textwidth}{!}{\includegraphics{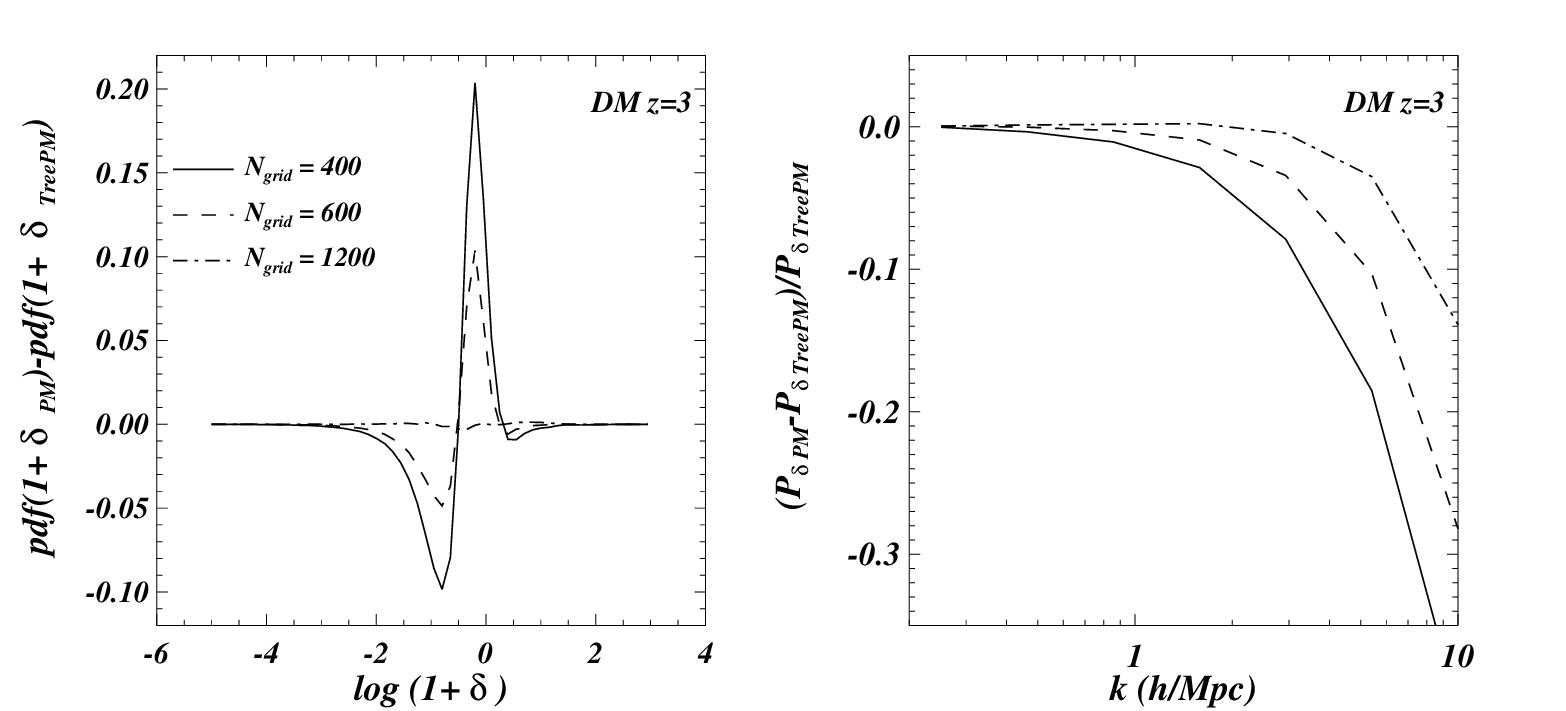}}
\caption{{\it Left:} differences in the probability distribution
functions of the dark matter density field between simulations run with
\gad in its HPM and in its TreePM mode (the PM grid for the TreePM run
is fixed to the value $N_{\rm grid}=200$). {\it Right:} Fractional
differences in the 3D matter power spectrum. The results are shown at
$z=3$ and for three different values of $N_{\rm grid}$ (400,600,1200)
as continuous, dashed and dot-dashed curves, respectively.}
\label{f4}
\end{figure*}

\begin{figure*}
\center\resizebox{1.0\textwidth}{!}{\includegraphics{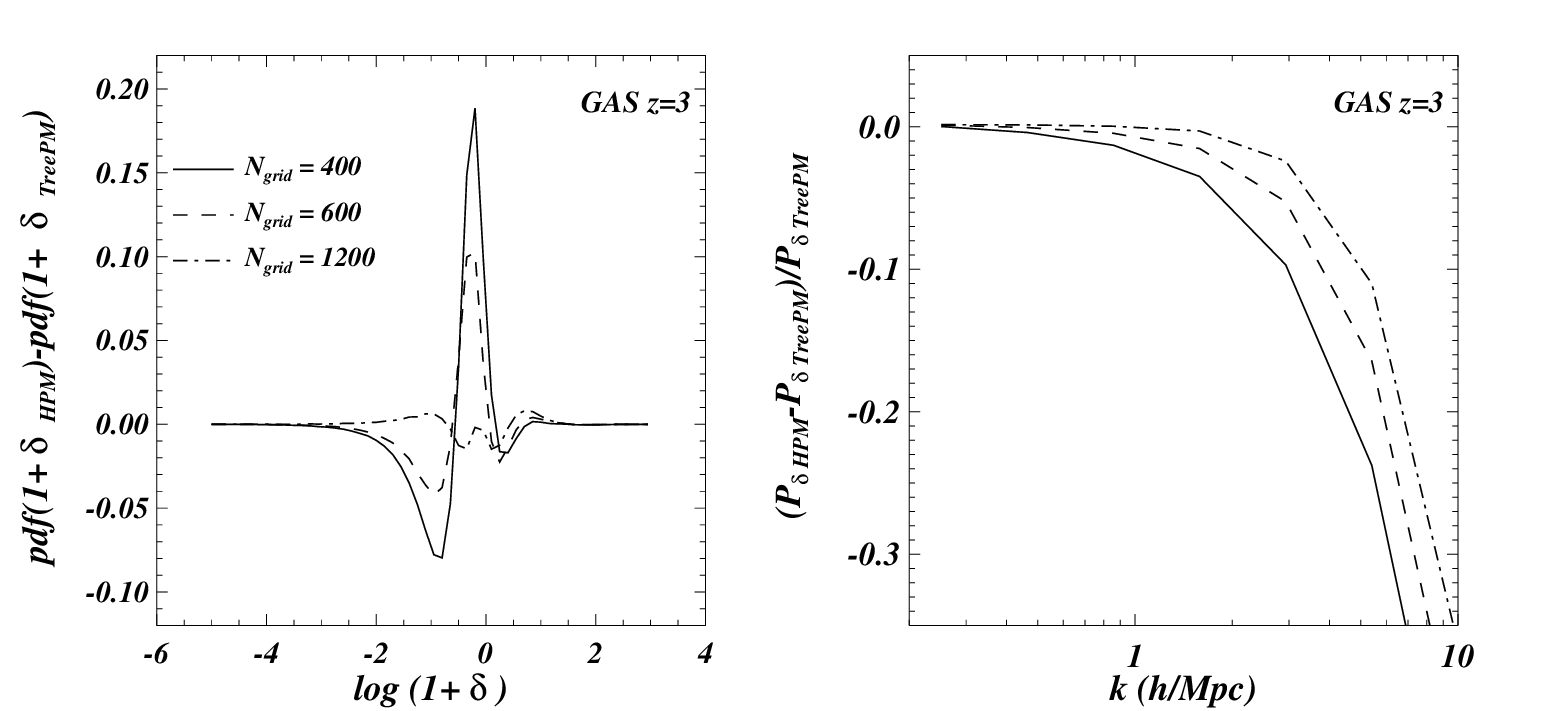}}
\caption{{\it Left:} differences in the probability distribution
functions of the gas density field between simulations run with \gad in
its HPM and in its TreePM mode (the PM grid for the TreePM run is fixed
to the value 200). {\it Right:} Fractional differences in the 3D matter
power spectrum. The results are shown at $z=3$ and for three different
values of $N_{\rm grid}$ (400,600,1200) as continuous, dashed and
dot-dashed curves, respectively.}
\label{f5}
\end{figure*}

We have implemented this HPM method in the simulation code {\small GADGET-2}.
We closely follow the approach of GH with only a few minor differences. In
the HPM code of GH, only one set of particles was used, i.e.~the fact that
the dark matter does not feel the pressure on small scales was neglected.  As
\gad is a SPH code which treats DM and baryons separately, we kept this
distinction in our HPM implementation. This may result in some small
differences on small scales. In Section \ref{compare1}, we will compare
simulations with the HPM implementation of \gad to runs carried out with the
HPM code of GH (kindly provided by Nick Gnedin).

There are three numerical parameters defining the technical details of
our HPM implementation in \gad. The first parameter is the number of
cells of the PM grid. We describe this by $N_{\rm grid}$, the number of
cells per dimension.  The second parameter, $H_{\rm s}$, describes the
scale of the smoothing applied to the enthalpy field before taking its
spatial derivative.  The density and enthalpy fields are more sensitive
to shot noise than the gravitational potential, because for the latter,
high frequency noise is suppressed as $\phi(k)\propto \delta_k \,
k^{-2}$.  We have thus followed GH and apply a Gaussian smoothing to
the density field before computing the enthalpy and its spatial
derivative. We apply a smoothing factor $\exp(-{k}^2 h_{\rm s}^2)$ to
the density field in Fourier space, where $h_{\rm s}= H_{\rm s}
L/N_{\rm grid}$. The third numerical parameter, $r_{\rm s}= A_{\rm s} L
/N_{\rm grid} $, is the scale of the smoothing of the PM force, which
we usually express in terms of $A_{\rm s}$, i.e.~in units of the mesh
cell size.  The parameter $A_{\rm s}$ hence controls the level of
residual force anisotropies in the PM force.  In the TreePM code,
$r_{\rm s}$ also gives the scale of the short-range/long-range force
split.  We will discuss the choice of numerical values for these
parameters in Section \ref{numparam}.  Note that the HPM code of GH has
only two parameters $N_{\rm grid}$ and $H_{\rm s}$, i.e.~no attempt is
made to make the PM force more isotropic on the scale of the mesh. GH
have adopted the choice $H_{\rm s}=3$.

To fix the slope and normalisation of the power-law temperature-density
relation of the IGM, our code follows the thermal history of two fiducial gas
elements at density values equal to the mean cosmic density and at 1.1 times
the mean cosmic density.  For a specified evolution of the ionising UV
background, we can then compute the values of $T_0$ and $\gamma$ from the
temperatures attained by these two fiducial gas elements.

In Figure \ref{f1}, we compare the 3D gas power spectrum for a range of HPM
simulations with different particle numbers and mesh sizes with a full
hydrodynamical simulation with $200^3$ dark matter and $200^3$ gas particles
(shown as triangles).  All simulations were run with \gad.  We only show
results at $z=3$, but note that the results at $z=2$ and $z=4$ are very
similar. On large scales ($k< 6 h$/Mpc), the power spectrum of the gas
distribution of HPM simulations converges nicely to that of the full
hydrodynamical simulation when the resolution of the mesh used for calculating
the gravitational forces is increased.  Note, however, that even for very high
resolution (six times more mesh cells in the HPM simulation than particles in
the full hydrodynamical simulation) the power on small scales in the HPM
simulations is significantly smaller than that in the full hydrodynamical
simulations.  Note further that changing the mesh resolution is more important
than changing the particle number in the HPM simulations. The thin and thick
solid curves are for HPM simulations with the same grid resolution but a
factor eight different particle number. They are virtually identical. We also
note that the results and trends for the dark matter power spectrum are
qualitatively similar.  In the runs discussed in the following, we will use the
HPM implementation of \gad with $2\times 200^3$ particles and with $N_{\rm
grid} \ge 200$.

\subsection{Comparison between the HPM implementation of \gad and the
HPM code of Gnedin \& Hui}
\label{compare1}

In this section we compare the gas and dark matter distribution of simulations
run with the HPM implementation of the \gad code and the HPM code of GH.  We
use the same initial conditions and temperature-density relation.
At $z=2$, $3$, and $4$, $T_0$ and $\gamma$
($T_0,\gamma$) have the following values: (21500 K,1.505), (21500
K,1.524) and (19200 K,1.536).

In Figures \ref{f2} and \ref{f3}, we show the relative differences of
the probability distribution and the power spectrum of the {\it dark
matter} and {\it gas} density at redshifts $z=2$, $z=3$, and $z=4$. We
have varied the resolution of the mesh to calculate the gravitational
force in the HPM implementation of \gad in steps of factors of two.
The other two relevant parameters in the \gad runs have been set to
$H_{\rm s} = 3$ and $A_{\rm s}= 1.25$.  For the case shown in Figures
\ref{f2} and \ref{f3}, the grid resolution for the HPM implementation
of \gad was a factor two higher than that used for the HPM code of
GH. In this case the agreement was best, better than 5\% (dark matter)
and 8\% (gas) for the probability distribution function{\footnote{The
pdf is defined as the number of points or pixels in a given x-axis bin with the
property that its integral along the x-coordinate is one.} (pdf) and better than 2\% for power spectra at
wavenumbers relevant for constraining cosmological parameters with the
\lya forest, $0.3\,\mincir k \, (h/{\rm Mpc}) \mincir 3$ (Viel,
Haehnelt \& Springel 2004). Because of the smoothing applied to the PM
force in \gad, a somewhat finer mesh is needed to match the results of
the HPM code by GH, where such a smoothing is not carried out and
larger force anisotropies on the mesh scale are accepted. By
reducing $A_{\rm s}$, the agreement of the two codes could be improved
further. The two HPM codes agree very well. In the following we will
only use the HPM implementation of \gad but our results should apply
similarly to the GH code.

\section{Comparison between full hydrodynamical and HPM simulations}
\label{compare2}

\subsection{The dark matter and gas density fields}
\label{dmgas}

\begin{table}
\begin{tabular}{llll}
\hline
\small (CODE, \# part.) & $N_{\rm grid}$ & CPU-time (ks) &  Mem. (Gb) \\
\hline

HPM-$200^3$ & 100 & 1.5 & 3  \\
HPM-$200^3$ & 200 & 3.1 & 3.5  \\
HPM-$200^3$ & 400 & 4.7 & 4.5 \\
HPM-$200^3$ & 600 & 11.2 & 12  \\
HPM-$200^3$ & 1200 & 15 & 76 \\
HPM-$400^3$ & 100 & 33 & 26 \\
HPM-$400^3$ & 200 & 35 & 28  \\
HPM-$400^3$ & 400 & 40 & 30 \\
HPM-$400^3$ & 600 & 44 & 36 \\
Hydro-$200^3$ & 200 & 183 & 3.2 \\
Hydro-$400^3$ & 400 & 11700 & 26 \\
GH HPM-$200^3$ & 200 & 5.4 & 3.5 \\
\hline
\end{tabular}
\caption{CPU-time required to reach $z=2$ for simulations of a 30 Mpc/h box
$\Lambda$CDM model and for several different resolutions and values of
the parameter $N_{\rm grid}$. The memory required is shown in the last
column.  All the values are wall-clock times for 32 CPUs (1.3 GHz
Itanium 2) of the SGI Altix 3700 (COSMOS) at DAMTP (Cambridge).}
\end{table}

We first want to check the agreement of the dark matter and gas
distributions between simulations run with the TreePM and HPM
implementations of \gad. In Figure \ref{f4}, we show the
differences in the density pdf and the power spectrum for the dark
matter distribution at $z=3$, for three different values of $N_{\rm
grid}$ used in the HPM simulation. The results at $z=2$ and $z=4$ are
similar. The simulations were run with $200^3$ and $2\times 200^3$
particles, respectively.  In the simulation with the Tree-PM
implementation, the number of mesh cells of the PM grid was set equal
to the number of particles. As also expected from the results shown in
Fig.~\ref{f1}, the differences become smaller with increasing
resolution of the PM grid used for the HPM implementation.  The
differences in the pdf of the DM density are smaller than 10\% (20\%)
for $N_{\rm grid}$=600 (400). If a very fine mesh of dimension $N_{\rm
grid}$=1200 is used, the pdf of the HPM simulation is indistinguishable
from that of the full hydrodynamical TreePM simulation.  For $N_{\rm
grid}$=600 (400) the discrepancy in the dark matter power spectrum
(right panel) is less than 2\% (4\%) for $0.2< k (h/{\rm Mpc}) <2$.
For $N_{\rm grid}$=1200, the difference is less than 0.5\% in the same
range of wavenumbers.  At larger wavenumber the differences in the
power spectra become much larger due to the much higher resolution
achieved with the TreePM code.  Note, however, that these small scales
are not used for the recovery of the dark matter power spectrum from
the \lya forest because of the uncertainties in the flux power spectrum
due to the thermal history and the metal contamination of the IGM (Kim
et al. 2004).

Figure \ref{f5} shows the difference in the gas distributions between
simulations with the HPM and TreePM implementations.  The differences are
similar to those found in the dark matter distribution.

\subsection{Flux statistics}
\label{fluxstats}
\subsubsection{The flux probability distribution function}

\begin{figure*}

\center\resizebox{1.0\textwidth}{!}{\includegraphics{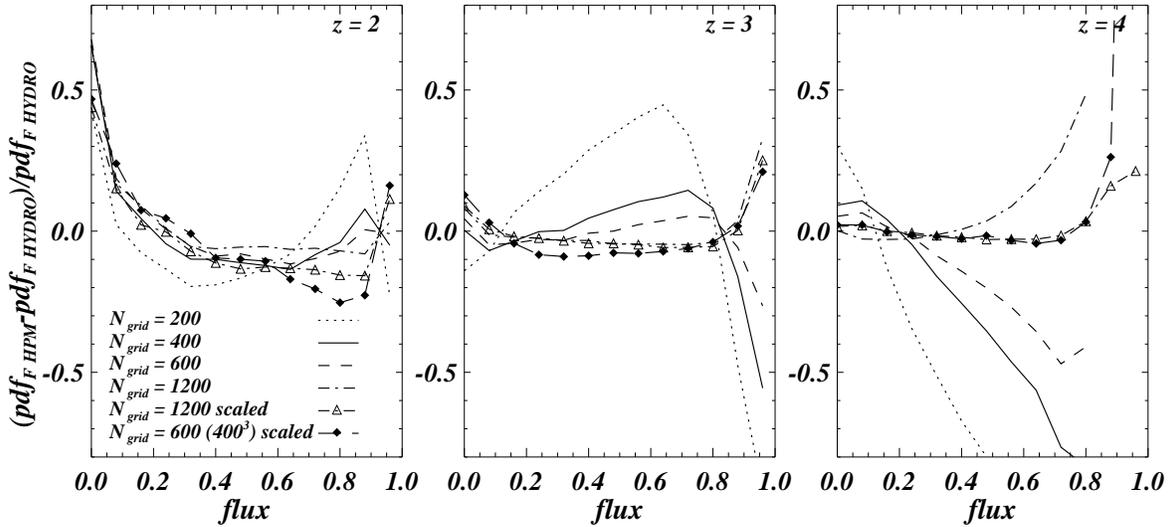}}
\caption{Effect of the parameter $N_{\rm grid}$ for simulations of a 30
Mpc$/h$ box with $2\times 200^3$ (gas and dm) particles. {\it Left
panel:} Fractional differences between the probability distribution
functions of simulations with $N_{\rm grid}$=200 (dotted), $N_{\rm
grid}$=400 (continuous), $N_{\rm grid}$=600 (dashed), and $N_{\rm
grid}$=1200 (dot-dashed) at $z=2$.  Also shown is the full
hydrodynamical TreePM simulation with $N_{\rm grid}$=200 and with the
same initial conditions. The long-dashed line with filled diamonds
represents results for the higher resoultion run with $2\times 400^3$
particles and with $N_{\rm grid}$=600.
The other HPM parameters have been fixed to
the fiducial values described in the text.  The dot-dashed curve with
overplotted empty triangles is for a simulation with $N_{\rm
grid}$=1200 for which the simulated flux has been scaled to match the
value of the full hydro simulations, in the other cases the spectra
have not been scaled (see text for the details).  {\it Middle Panel:}
Results at $z=3$.  {\it Right Panel:} Results at $z=4$.  }
\label{f6}
\end{figure*}

The flux distribution in the \lya forest depends on the spatial distribution,
the peculiar velocity field and the thermal properties of the gas. In the last
section, we have shown that the gas distribution of the HPM simulations
converges rather well to that of the full hydrodynamical simulations when the
resolution of the PM mesh is improved.  For the flux distribution the
situation is more complicated, however. In Figure \ref{f6}, we plot the
differences in the pdf of the flux for HPM simulations with a range of $N_{\rm
  grid}$ values compared with the full hydrodynamical simulations at
$z=2,3,4$.  The simulations are the same as those discussed in section
\ref{compare2} and shown in figures \ref{f4} and \ref{f5} (these figures
show results only at $z=3$).  The curves without symbols show the results
for the same amplitude of the ionising UV background as in the full
hydrodynamical simulations.  Note that this means that the the flux
distribution has {\it not} been rescaled to a fixed mean flux, as it is often
done.  Such a rescaling would mask the numerical effects we seek to identify
here. However, to facilitate comparison with other work (e.g. ~McDonald et al.
~2004a), the curves with triangles show the pdf of the flux after re-scaling
the flux distribution of the $N_{\rm grid}=1200$ HPM simulation such that the
mean transmitted flux is the same as in the full hydrodynamical simulations.

At $z=3$, the flux distribution of the HPM simulations converges reasonably
well to that of the full hydrodynamical simulations. With the exception of
flux levels $F>0.8$, the differences are smaller than 5\% for $N_{\rm
  grid}=600$ and even smaller for higher resolutions of the PM mesh.  In
regions of low absorption ($F> 0.8$) the differences are, however,  large
(10-20\%), change sign with increasing resolution, and do not converge. We
have inspected a few spectra individually and found that the discrepancy is
due to differences in both density and temperature in the lowest density
regions.  At $z=4$ these differences in regions of low absorption are
substantially larger.  Because of the strong decrease of the mean flux with
increasing redshift, these regions correspond to significantly more underdense
regions than at $z=3$. At $z=2$ additional large differences up to 50\% arise
in regions of strong absorption, which also do not vanish with increasing
resolution. For the $N_{\rm grid}=1200$ HPM simulation, the overall agreement
with the full hydrodynamical simulation is of the order of 2\% for $F<0.85$ at
$z=3,4$, while at $z=2$, discrepancies of the order of $\magcir 10$\% remain
both in underdense and very dense regions.  The differences at $z=2$ and for
$F<0.15$ are due to the gas in dense regions being substantially colder in the
HPM simulations than in the full hydrodynamical simulation where a significant
portion of the dense gas is shock heated. In Figure 6 we overplot the
results from a higher resolution HPM run with $2\times 400^3$ particles
and $N_{\rm grid}=600$ as a long dashed line with filled diamonds. The
results are very similar to the HPM simulation with $2\times 200^3$
particles and $N_{\rm grid}=200$.
We hence confirm the findings of
GH that the differences in the flux pdf between HPM and full hydrodynamical
simulations are of the order of 10-15\%.

\subsubsection{The flux power spectrum}

The main motivation of the use of HPM simulations comes presently from the
need for accurate predictions of the flux power spectrum for a wide range of
astrophysical and cosmological parameters. Such a grid of predictions allows a
detailed comparison with observational data and a determination of best-fit
values and confidence intervals of cosmological parameters (McDonald et
al. 2004a). 

In Figure \ref{f7}, we plot the differences of the flux power spectrum of
HPM simulations with a range of mesh sizes compared with full hydrodynamical
simulations at $z=2,3,4$.  The simulations are the same as those discussed in
the previous sections.  As in figure \ref{f6}, the curves without symbols
show results for the same amplitude of the ionising UV background while
the curves
with empty triangles show the flux power spectrum after rescaling the flux
distribution of the $N_{\rm grid}=1200$ HPM simulation such that the mean flux
is the same as in the full hydrodynamical simulations. In Figure
\ref{f7} we show the results from a higher resolution HPM run with
$2\times 400^3$ particles and  $N_{\rm grid}=600$, as the long-dashed
line with overplotted filled diamonds. At redshift $z=4$ and $z=3$
there is perfect agreement with the  $N_{\rm grid}=1200$ HPM
simulation, in the wave number range of interest here. At $z=2$ there
are small differences of the order of $< 5\%$. Thereby, increasing the
number of particles does not improve the agreement significantly.

At redshifts $z=3$ and $z=4$, the flux power spectra of the HPM
simulations converge well to those of the full hydrodynamical
simulations, but only for resolutions of the PM mesh where the number
of the mesh cells is substantially larger than that of the number of
particles in the full hydrodynamical simulations. At $z=3$, the HPM
simulations with $N_{\rm grid}$=400 (600) have {\it scale-dependent}
differences of about 10\% (7\%) in the wavenumber range relevant for
inferring the matter power spectrum.  For $N_{\rm grid}$=1200, there is
a scale-independent offset of about 5\% (3\% when rescaled to the same
mean flux). At redshift $z=4$ the situation is very similar. However,
at redshift $z=2$, the flux power spectrum of the HPM simulations does
not converge to that of the full hydrodynamical simulation. The
differences are here actually smallest for the HPM simulation with
lowest resolution ($N_{\rm grid}$=200).  However, even in this case the
discrepancies are large and strongly scale dependent, of the order of
25-30\% at the largest scales.  At small scales $k > 0.02$ s/km, the
size of the disagreement and its scale dependence is similar to that
found by McDonald et al. (2004a, their figure 5).  Note that because of
the smaller box size of their hydro simulations, McDonald et al.~were
not able to probe scales $k < 0.007$ s/km (at $z=2$), where the
differences increase dramatically.  Note that the amount of
shock-heated gas is significantly larger in simulations with larger
box-size. To test further to what extent these discrepancies at large
scales depend on the resolution of the hydro-simulation, we have run an
additional hydro-simulation with 64 times higher mass resolution
($2\times 200^3$ particles in a 7.5 $h/$ Mpc box). There is good
agreement with the results shown in Figure 7.  We stress here
that our goal is to get a good convergence of the flux power in the
range $0.003 < k$ (s/km) $<0.03$, which is the range which is used for
the matter power spectrum reconstruction as in Viel et al. (2004).

These large differences and the lack of convergence appear
perhaps counterintuitive considering the rather good convergence of the gas and dark
matter distribution.  However, they simply originate in the large differences
in the pdf of the flux distribution, which in turn are due to the different
thermal state of the gas in high density regions in the HPM and full
hydrodynamical simulations.

At redshift $z=2$, a larger proportion of the
absorption is from gas in high density regions, which is shock-heated in the
full hydrodynamical simulations and therefore on average hotter than in the
HPM simulations.  This tends to mainly affect the strong absorption systems
which contribute significantly to the flux power spectrum at large scales
(Viel et al. 2004a; McDonald et al.  2004b). We fill discuss this further in
the next section.

\begin{figure*}
\center\resizebox{1.0\textwidth}{!}{\includegraphics{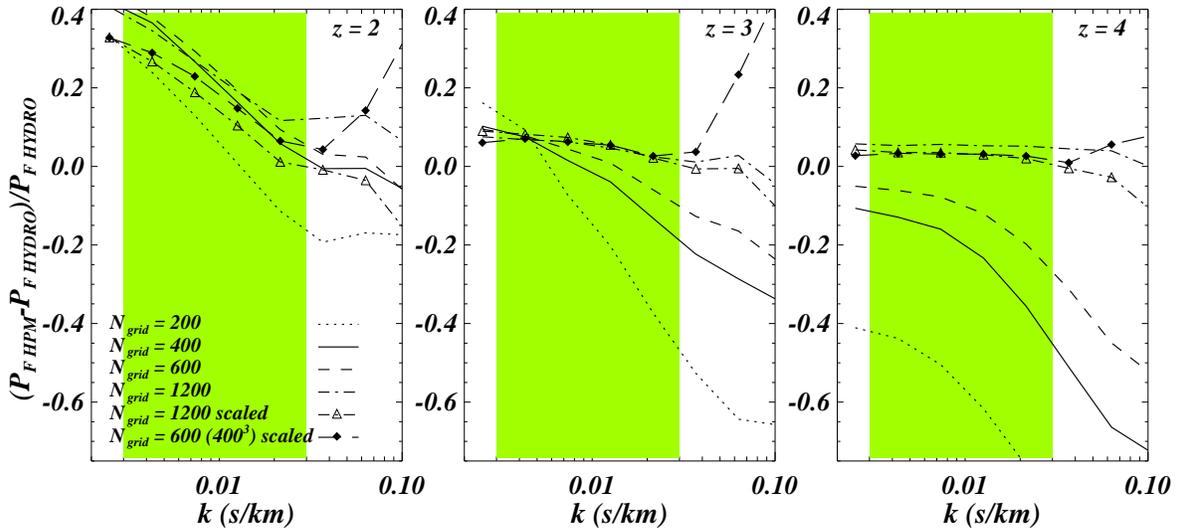}}
\caption{Effect of the parameter $N_{\rm grid}$ for simulations of a 30
Mpc$/h$ box with $2\times 200^3$ (gas and dm) particles. {\it Left
panel:} Fractional differences between the 1D flux power spectra of
simulations with $N_{\rm grid}$=200 (dotted), $N_{\rm grid}$=400
(continuous), $N_{\rm grid}$=600 (dashed), and $N_{\rm grid}$=1200
(dot-dashed) at $z=2$. Also shown is the full hydrodynamical TreePM
simulation with $N_{\rm grid}$=200 and the same initial conditions. The
other HPM parameters have been fixed to the fiducial values described
in the text.  The dot-dashed curve with overplotted empty triangles is
for a simulation with $N_{\rm grid}$=1200 for which the simulated flux
has been scaled to match the value of the full hydro simulations, in
the other cases the spectra have not been scaled (see text for the
details).  The long-dashed line with filled diamonds represents results
for the higher resoultion run with $2\times 400^3$ particles and with
$N_{\rm grid}$=600 (results scaled to reproduce the same $\tau_{\rm eff}$).
{\it Middle Panel:} Results at $z=3$.  {\it Right
Panel:} Results at $z=4$.  In all the panels the dashed area represents
the range of wavenumbers used by Viel, Haehnelt \& Springel (2004) to
recover cosmological parameters.  }
\label{f7}
\end{figure*}

\subsubsection{Temperature effects  on the flux pdf and the
flux power spectrum}
\label{shock}

We have argued that the approximation of the  relation between
gas density and gas temperature as a power-law breaks down at low redshift.  This
approximation inevitably does not take into account the amount of moderately
shock-heated gas that is falling into the potential wells of the dark matter
haloes.  In this section, we want to check this  explicitly.  
For this purpose we use the hydrodynamical simulation and the HPM
run with $N_{\rm grid}$=600.

As a first step we perform the following simple test. We superimpose
onto the full hydro-dynamical SPH simulation the temperature-density
relation of the HPM runs, and then recompute the QSO spectra. We find
that this results in differences much smaller than those in Figure 7,
of order 8\%, 5\% and 4\%, at $z=2$, $3$ and $4$, respectively, at the
largest scales.  Most of the discrepancy is thus indeed due to the
differences in the thermal state especially at low
redshift. Differences in the thermal state will lead, however, also to
pressure differences during the dynamical evolution, which will modify
the mass distribution and the peculiar velocity field. In shock fronts,
the change in particle trajectories can be substantial. Since the HPM
implementation does not capture shocks, it would not treat the dynamics
correctly even if the temperatures would be accurate at all times. To
investigate this further we have run an SPH simulation with artificial
viscosity set to zero and the temperature-density relation of the HPM
simulation. This should mimick an `ideal' HPM simulation: the
gravitational force is resolved with high accuracy and in an isotropic
way, while the pressure gradients are smooth and resolved everywhere
with the maximum resolution allowed by the local particle sampling. The
standard HPM method has a less well resolved gravitational force and
should be sensitive to over- or under-smoothing of the pressure field
in regions of high or low particle density, respectively.  The results
are shown in Figure \ref{f8}. The SPH simulation is represented by
the dashed line while the dotted line is for the HPM simulation with
$N_{\rm grid}=1200$ (both the runs have the same number of particles
equal to $2\times 200^3$).  There is good agreement with the HPM
simulations, suggesting that the discrepancy in the flux power is
primarily due to the different thermal state of the gas due to shocks
and not to any artefacts of our particular HPM implementation.  The
total effect on the flux power spectrum should thereby be a combination
of an increase of the overall amount of shock-heated gas with
decreasing redshift and the change of the mean effective optical depth.
The flux power spectrum becomes increasingly sensitive to
higher-density gas with decreasing redshift due to the decreasing
effective optical depth.

We will now investigate the relation between the differences  between HPM and
SPH and gradients in the velocity field of the gas.  Negative gradients in 
the peculiar velocity field along the line-of-sight should 
represent a signature of infalling material and  may thus serve as 
a rough guide to where shocks occur. 
In the left panel of Figure~\ref{f9}, we show the ratio of the gas temperatures for 
the full hydrodynamical simulation and the HPM simulation 
as a fuction of the velocity gradient of the gas. 
We first average the temperature in  pixels within 100
km/s from a minimum in the gradient of the peculiar velocity field in
real space.  Then, we average over the corresponding flux values, in
redshift space.  Before selecting the negative gradients in the
hydrodynamical simulations, we have explicitely checked that the
peculiar velocity fields are very similar in both simulations.

\begin{figure*}
\center\resizebox{1.0\textwidth}{!}{\includegraphics{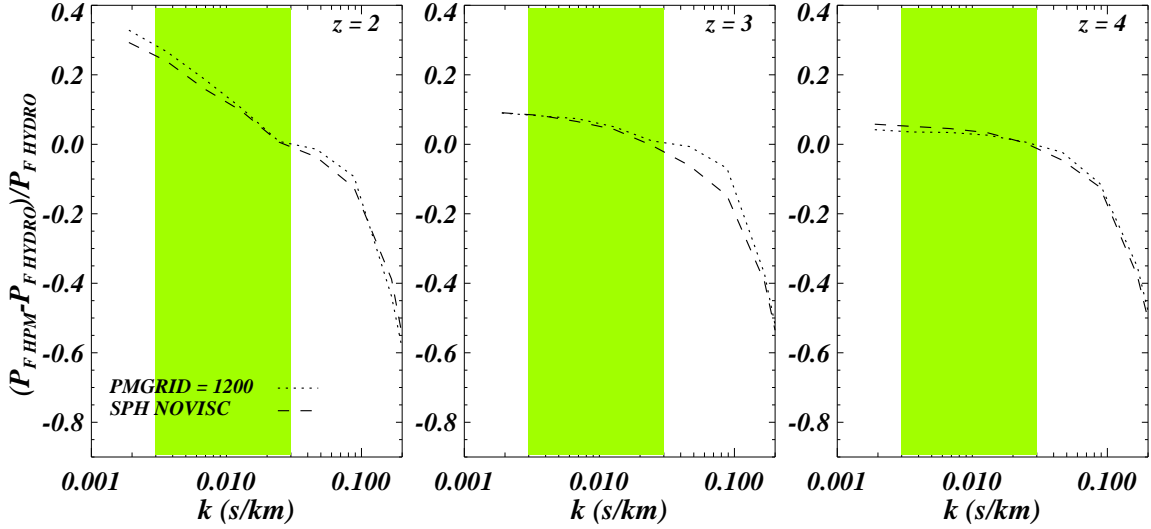}}
\caption{Fractional differences between: 1) an HPM simulation with $2\times
200^3$ and with $N_{\rm grid}=1200$ and a full SPH hydrodynamical
simulation (dotted line); 2) between a SPH  simulation with zero artificial
viscosity and with a superimposed temperature-density relation of the
HPM runs and a full SPH hydrodynamical simulation (dashed
line). Results are shown at $z=2,3,4$ in the left, middle and right
panels, respectively. Spectra have been scaled to reproduce the same
effective optical depth.}
\label{f8}
\end{figure*}

\begin{figure*}
\center\resizebox{1.0\textwidth}{!}{\includegraphics{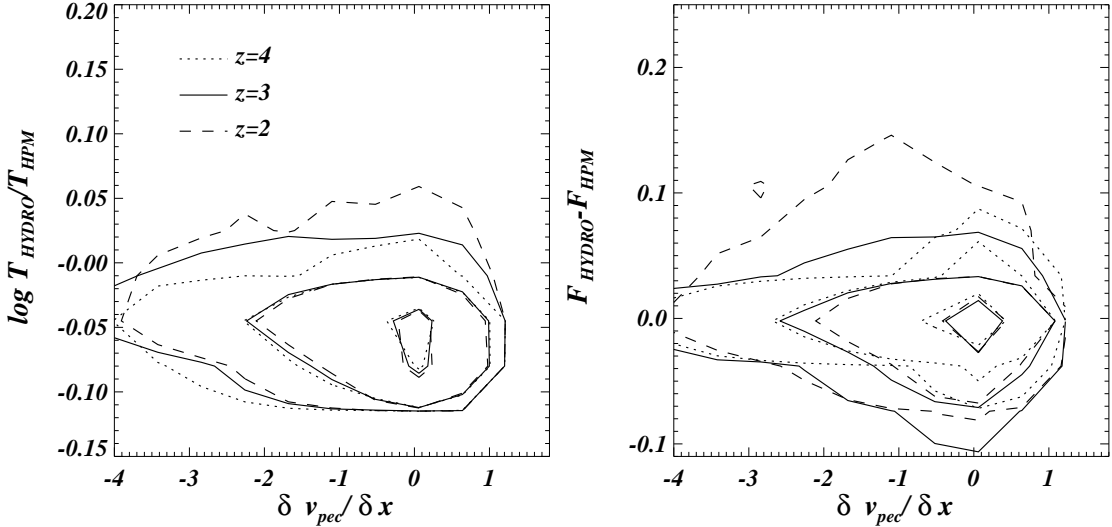}}
\caption{Role of shock-heated gas. {\it Left:} Ratio of the
temperatures in simulations run with the full hydro and HPM
implementation with $N_{\rm grid}$=600 with the same initial
conditions, plotted as a function of the gradient of the peculiar
velocity field along the line of sight. {\it Right:} Differences in the
simulated flux values.  The contour plots represent the number density
of points in the 2D plane and the number density increases by an order
of magnitude at each contour level. The dashed, continuous and dotted
lines are for $z=2,3,4$, respectively.}
\label{f9}
\end{figure*}

The contour plots indicate the number density of points, which varies
by a factor of 10 between adjacent contour levels. The bulk of the pixels in
this panel is in regions with $\delta v/\delta x \sim -0.5$ and at
`hydrodynamical' temperatures that are about 10\% lower than the corresponding
temperatures of the HPM simulation.  The simulation at  $z=2$  shows a
significantly increased amount of pixels at $\delta v/\delta x \sim -1$ with
HPM temperatures that are colder than the corresponding temperatures in the
hydrodynamical simulation.  These differences in the temperatures have an
important effect on the simulated flux. 

In the right panel of Figure~\ref{f9}, we plot the differences in the flux for
the same regions of infalling gas.  Since the flux is observed in redshift
space we have  averaged the flux within 100 km/s velocity bins.  There
are no obvious trends for smaller values of $\delta v/\delta x$ (i.e.
``stronger'' shocks). This is due to the fact that stronger shocks have a more
complex temperature and density structure which in the hydro simulation is
represented more faithfully than in the HPM simulation. As a result, the
differences in the temperatures and fluxes actually tend to be averaged out
for strong shocks.  The scatter for positive values of the gradient $\delta
v/\delta x$ also shows smaller scatter, both in the temperatures ratio
and in the flux differences.  Most of the differences at $z=2$ arise from
regions of infalling gas that are not modelled accurately by the HPM method.
This suggests  that at least part of the discrepancy at low
redshift is due to the increased amount of shock-heated gas probed by
the Ly-$\alpha$ forest at lower redshift.

\subsection{The effect of the numerical parameters $H_{\rm s}$ and $A_{\rm s}$ }  
\label{numparam}

As discussed in section \ref{hpm}, we need to specify the parameters $H_{\rm
  s}$ and $A_{\rm s}$ which describe the smoothing of the gas density and of
the gravitational force field in the HPM simulations.  There is no obvious
optimum choice for these parameters, so a  choice needs to be made by
comparing to the full hydrodynamical simulations.  For changes of $H_{\rm s}$,
which controls the smoothing of the pressure field, the resulting differences
are very small at large scales (less than 1\%).  They are only weakly scale-
and redshift-dependent and only slightly increase at small scales for $H_{\rm
  s}$ in the range 1.5-3. We have therefore fixed $H_{\rm s} = 3$ for all
simulations.  

Varying the parameter $A_{\rm s}$, which controls the smoothing of the
gravitational force field, has a somewhat larger effect.  In Figure
\ref{f10}, we show the differences between the flux power spectrum of a HPM
($N_{\rm grid}=600$) simulation and that of the full hydrodynamical simulation
for different values of $A_{\rm s}$, at three different redshifts $z=2,3,4$.
The differences are typically a few percent but can be as large as 10\%. It is
not obvious which value for $A_{\rm s}$ represents an optimum choice.  One
possibility is to impose a certain requirement for the maximum allowed force
anisotropy generated by a point mass in the PM scheme. Using such a criterion,
we have set $A_{\rm s}= 1.25$, which gives typical PM force errors less than 1
per cent.

\begin{figure*}
\center\resizebox{0.5\textwidth}{!}{\includegraphics{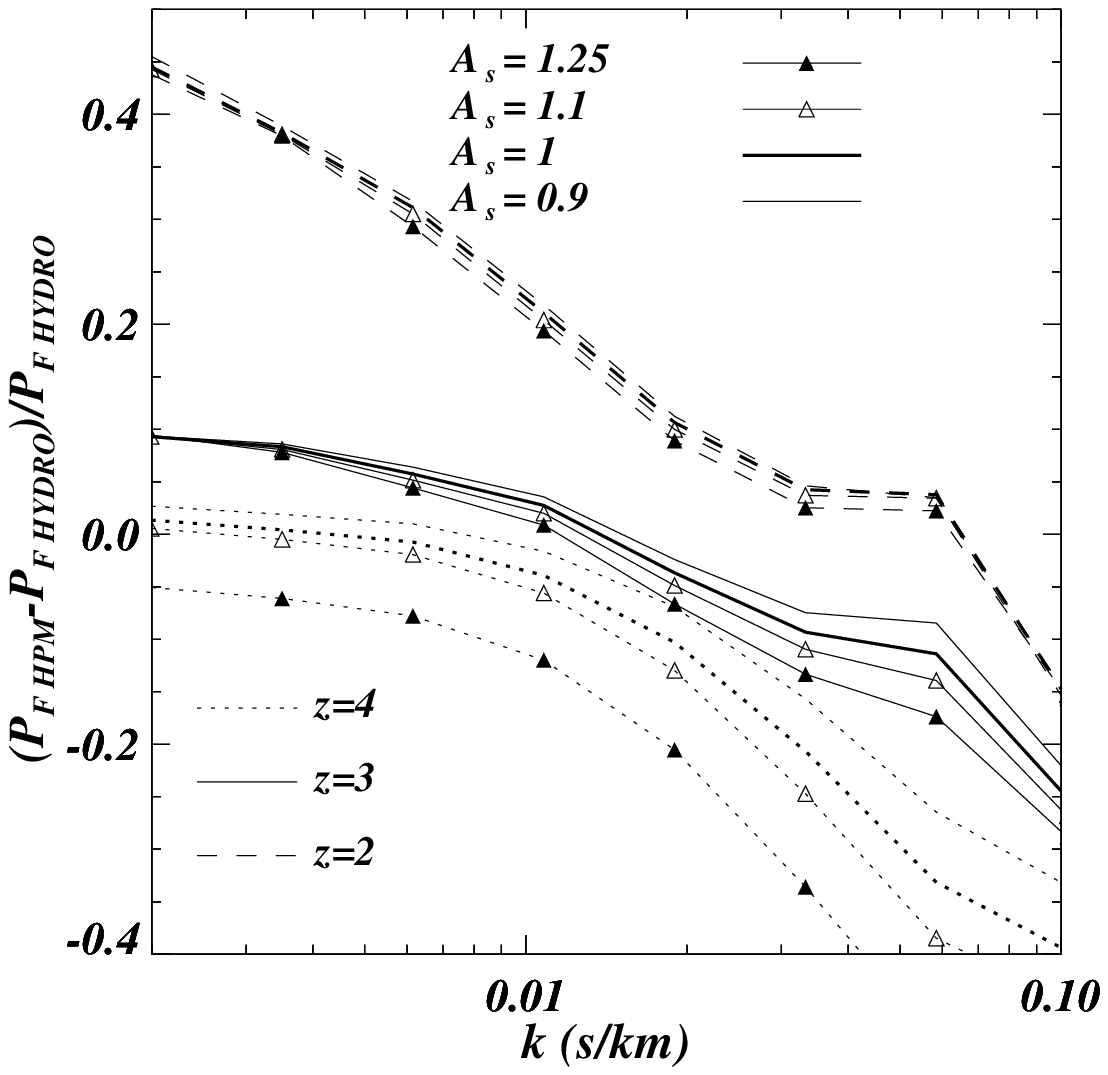}}
\caption{Effect of the smoothing parameter $A_{\rm s}$ of the HPM
implementation in $\gad$ for simulations of a 30 Mpc$/h$ box with
$2\times 200^3$ particles. We plot the fractional differences between
the 1D flux power spectra of a model with $A_{\rm s}$ set to 1.25 (thin
line with filled triangles), 1.1 (thin line with empty triangles), 1
(thick line), 0.9 (thin line) and the full hydrodynamical simulation.
The results are shown for three different redshifts $z=2,3,4$ as 
dashed, continuous and dotted curve, respectively.  Note that the flux
power spectra have not been scaled to reproduce the same effective optical
depth.}
\label{f10}
\end{figure*}

\subsection{CPU time and memory requirements}
\label{cputime}
In Table 1, we summarise the total CPU time (in wall-clock seconds) required
by the simulations to run to $z=2$, and their memory requirement (in Gbytes).
We include simulations with a range of particle numbers and resolutions of the
PM mesh, all for a box size of 30 Mpc$/h$.  The HPM simulation with $N_{\rm
  grid}$=600 has run about 20 times faster than the hydrodynamical SPH/TreePM
simulation at the corresponding resolution, but has a three times larger
memory requirement.  The $N_{\rm grid}$=1200 HPM simulation, which as we saw
gave a good agreement with the full hydrodynamical simulations in terms of the
gas and dark matter distribution, is still faster than the SPH simulation by a
factor 10, but its memory requirement is very large. We note that the
simulations with a very high resolution of the PM mesh ($N_{\rm grid} = 1200$)
have been difficult to run because of their very large memory requirement,
which is close to the total amount available on the COSMOS computer we used.

\section{Discussion and Conclusions}
\label{conclu}

We have compared full hydrodynamical simulations carried out with the
SPH/TreePM code \gad to simulations that used the HPM method. The latter
scheme was implemented by us in \gad, and we compared this implementation with
the independent code of GH.  Our comparison was performed at redshifts
$z=2$,$\,z=3$ and $z=4$. Our main results can be summarised as follows.

\begin{itemize}   
  
\item{The dark matter and gas distributions of HPM simulations with \gad
    converge well to the full hydrodynamical simulations with the SPH/TreePM
    code.  For a PM mesh with $>6^3$ more mesh cells than particles in the SPH
    simulations, the difference in the pdf of the gas and matter distributions
    are less than 1 percent.  The same is true for the matter power spectrum
    at wavenumbers up to 20 times the fundamental mode of the box for a mesh
    with $1200^3$. At smaller scales the differences in the power spectra
    strongly increase due to lack of resolution of the HPM grid.}
 
\item{The pdf of the flux distribution of HPM simulations with \gad does not
    converge to that of the full hydro simulations.  At low levels of
    absorption ($F>0.8$) the differences (10\% and more) do not decrease with
    increasing resolution at all three redshifts examined.  At $z=2$, there is
    an additional large difference at low flux levels which rises to 50\% at
    the lowest flux levels. The latter difference is most likely due to the
    larger proportion of absorption by dense shock-heated gas at $z=2$ which
    is not modelled well by the HPM method. }

\item{At redshifts $z=3$ and $z=4$, the flux power spectrum of HPM simulations
    with \gad does converge to that of the full hydrodynamical simulations up
    to a scale independent offset.  For a HPM simulation with box size of
    30\mpch and a PM mesh with $1200^3$ cells this offset is about $5\% -7\% $
    at wave numbers 0.002 s/km $<k<$ 0.05 s/km.  At $z=2$, however, there are
    large scale-dependent differences between the flux power spectrum of the
    HPM simulation and the full hydrodynamical simulation which are as large
    as 20-40\%.  These differences are again most likely due to the larger
    proportion of absorption by dense shock-heated gas at $z=2$.}

\item{The HPM implementation of \gad and the code by GH give similar results
    (to within a few percent) for the same initial conditions, provided a
    slightly higher resolution of the PM grid is used for \gad. This
    offset is a result of the PM force smoothing done by \gad, which is
    adjustable.  The results obtained above should thus hold in a similar
    form for the GH code. }

\end{itemize}

 The HPM method involves two main simplifications compared to full
hydro-simulations, calculating the pressure in an approximate way and
estimating the temperatures based on the density alone.  The HPM
approximation does a good job in modelling the gas and matter
distribution on the scales relevant for the Lyman-$\alpha$ forest
suggesting that the first approximation works well.  The situation for
an accurate prediction of the flux distribution is quite different and
we have shown that the treatment of the thermal state in the HPM
approximation is the main problem for accurate predictions of the flux
distribution.  The strong dependence of the transmitted flux on the
thermal state of the gas together with the crude approximation of the
thermal state in the HPM approximation leads to large and not always
intuitive scale- and redshift-dependent differences in the flux
distribution between HPM and the full hydrodynamical simulations.

For the flux power spectrum, these
differences are less important than for the pdf of the flux distribution. Our
results suggest that at $z=3$ and $z=4$ the gain in speed offered by HPM
simulations may still make them an attractive tool to obtain predictions of
the flux power spectrum for a wide range of parameters. This will, however,
require very careful calibration with full hydrodynamical simulations, and it
appears doubtful that HPM simulations are suitable to model the dependence of
the flux power spectrum on the thermal state of the gas accurately. The rather
large memory requirement of HPM simulations with sufficient resolution to
reach convergence also partially offsets the advantage of their higher speed.
Our results further suggest that at lower redshift the larger proportion of
absorption by dense shock-heated gas makes HPM simulations unsuitable for
accurate predictions of the flux power spectrum.

	  Currently the observational uncertainties regarding the
	  thermal state of the IGM are still rather large. The results 
	  of quantitative studies of the matter power spectrum with Lyman-alpha
	  forest data are therefore generally marginalized over a wide 
     range of simple temperature-density relations. The
     difficulties of simple HPM implementations with modeling the
     effect of the thermal state accurately may therefore    
     be less important than suggested by our discucussion so far. 
     However, improved measurements  of the thermal
     state of the gas utilizing the Doppler parameter distribution, 
     the flux PDF and the small scale flux power spectrum 
     are an important prerequisite for reducing the errors of
     measurements of the matter power spectrum from \lya forest data.  
	
     Accurate modeling of the thermal state of the gas will be
     required to take full advantage of an reduced uncertainty
     regarding the thermal state of the IGM. For HPM simulations
     this will almost certainly require a signifificant improvement
     of the modelling of the thermal state, e.g. by introducing some
     scatter in the temperature density relation. Full
     hydrodynamical simulations could thereby be used to quantify
     and calibrate this scatter and to investigate possible
     correlations of the scatter with physical quantities. Such
     modelling would obviously greatly benefit from more precise
     observational estimates of the parameters describing the
     temperature-density relation which may be possible with the use
     of the flux power at smaller scales and from an estimate of the
     scatter in the temperature density relation using higher order
     statistics such as the bispectrum (Mandelbaum et al. 2003, Viel
     et al. 2004, Fang \& White 2004). It will then also be
     important (in HPM and full hydro simulations) to model other
     physical aspects affecting the thermal state of the gas as the
     presence of galactic winds and temperature/UV fluctuations due
     to the reionization of HeII.

\section*{Acknowledgements.} 
The simulations were run on the COSMOS (SGI Altix 3700) supercomputer
at the Department of Applied Mathematics and Theoretical Physics in
Cambridge and on the Sun Linux cluster at the Institute of Astronomy in
Cambridge. COSMOS is a UK-CCC facility which is supported by HEFCE and
PPARC. MV thanks PPARC for financial support and Adam Lidz for useful
discussions. MV, MGH and VS thank the Kavli Institute for Theoretical
Physics in Santa Barbara, where part of this work was done, for
hospitality during the workshop on ``Galaxy-Intergalactic Medium
Interactions''. This work is partly supported by the European Community
Research and Training Network ``The Physics of the Intergalactic
Medium''. We thank Nick Gnedin for providing us with a copy of
his HPM code.

\end{document}